\documentclass[12pt]{article}

\RequirePackage{latexsym}
\usepackage[]{graphicx}

\usepackage{underscore}
\usepackage{natbib}
\usepackage{amsopn}
\usepackage{subeqn}
\usepackage{subfigure}
\usepackage{mathptmx}
\usepackage[left=1in,top=1in,right=1in,bottom=1in]{geometry}

\newcommand{\beql}[1]{\begin{equation} \label{#1}}
\newcommand{\beq}{\begin{equation}}
\newcommand{\eeq}{\end{equation}}
\newcommand{\bea}{\begin{eqnarray}}
\newcommand{\eea}{\end{eqnarray}}
\newcommand{\bes}{\begin{subequations}\begin{eqnarray}}
\newcommand{\ees}{\end{eqnarray}\end{subequations}}
\newcommand{\mrm}[1]{\mathrm{#1}}
\newcommand{\mbf}[1]{\mathbf{#1}}
\newcommand{\msf}[1]{\mathsf{#1}}
\newcommand{\abs}[1]{{\vert {#1} \vert}}
\DeclareMathOperator{\sech}{sech}
\DeclareMathOperator{\eps}{eps}
\newcommand{\epsof}[1]{\eps({#1})}
\DeclareMathOperator{\sqrtfun}{sqrt}
\DeclareMathOperator{\gkintfun}{gkint}
\newcommand{\gkintfunof}[1]{\gkintfun({#1})}
\DeclareMathOperator{\absfun}{abs}
\newcommand{\absfunof}[1]{\absfun({#1})}

\usepackage[ruled]{algorithm2e}

\SetAlFnt{\small}
\SetAlCapFnt{\small}
\SetAlCapNameFnt{\small}
\SetAlCapHSkip{0pt}
\IncMargin{-\parindent}

\usepackage{fancyhdr}
\begin{document}

\title{AMGKQ: An Efficient Implementation of Adaptive Multivariate Gauss-Kronrod Quadrature for Simultaneous Integrands in Octave/MATLAB}
\date{October 4, 2014}
\author{Robert W. Johnson \thanks{\href{mailto:robjohnson@alphawaveresearch.com}{robjohnson@alphawaveresearch.com}}\\
\small Alphawave Research, Jonesboro, GA 30238, USA\\}






\maketitle

\begin{abstract}
The algorithm AMGKQ for adaptive multivariate Gauss-Kronrod quadrature over hyper-rectangular regions of arbitrary dimensionality is proposed and implemented in Octave/MATLAB.  It can approximate numerically any number of integrals over a common domain simultaneously.  Improper integrals are addressed through singularity weakening coordinate transformations.  Internal singularities are addressed through the use of breakpoints.  Its accuracy performance is investigated thoroughly, and its running time is compared to other commonly available routines in two and three dimensions.  Its running time can be several orders of magnitude faster than recursively called quadrature routines.  Its performance is limited only by the memory structure of its operating environment.  Included with the software are numerous examples of its invocation.
\end{abstract}

\section{Introduction}

As \citet{Press-1992} state:
\begin{quote}
``Integrals of functions of several variables, over regions with dimension greater than one, are \textit{not easy}.''
\end{quote}
We aim to make it so, at least for regions given by a hyper-rectangular volume in an arbitrary number of dimensions.  Since one often requires the integration of several functions over the same region, an algorithm that performs the evaluations simultaneously is inherently more efficient than repeating the quadrature independently for each function.  The use of a vectorized computing language, such as Octave or MATLAB, allows one to implement these evaluations with a minimum of coding.

In this article we propose AMGKQ, abbreviation for Adaptive Multivariate Gauss-Kronrod Quadrature. The main
contributions of this work can be summarized as follows:
\begin{itemize}
\item Vector, matrix, and binary singleton expansion operations are leveraged for efficiency.
\item Variable transformations are used for improper integrals in multiple dimensions.
\item An arbitrary number of breakpoints in multiple dimensions is possible.
\end{itemize}
The implementation AMGKQ.M is based on ADAPT.M by Alan Genz \citep{vanDooren-1976207,Genz-1980295,Berntsen-1991437} and QUADGK.M by David Bateman \citep{Octave-2009}, and it makes use of subroutines provided by Walter Gautschi [\citeyear{Gautschi-199421,Gautschi-2004}] and Randall LeVeque [\citeyear{LeVeque-2007}].  The advice given by \citet{Shampine-2008131} ``to vectorize the evaluation of functions'' is taken to extremes, as all the integrand values for each subregion are evaluated with a single call to the user supplied function.  The quadrature coefficients are stored in persistent variables to reduce the computational load, and care is taken throughout to compute what is needed only once if possible.  In the quest for efficiency, every FLOP counts.

The intended application of AMGKQ is in the context of Bayesian data analysis, where one usually finds a strongly peaked evidence density somewhere on the coordinate manifold against which expectation values of the observables are taken, but it should meet the requirements of a general purpose algorithm on par with those provided by Octave and MATLAB.  Standard variable transformations are used when an improper integral is detected \citep{Shampine-2010195}, and internal singularities are avoided by use of breakpoints defining boundaries within the integration region.  For good measure, it also does complex line (contour) integrals in the complex plane using the same machinery.

\section{Statement of the Algorithm}

In this section we will state the requirements of the algorithm, its initialization, and its main control loop.  The basic theory of Gauss-Kronrod quadrature is assumed to be known to the reader, as are the implementations ADAPT.M and QUADGK.M.  How the algorithm is used for contour integrals will be described at the end of this section.

\subsection{Definition of the Integrand, the Region, and the Initial Subregions}

The form of the user supplied function(s) $F_f (x,y,\ldots)$ is important to the efficient implementation of its integration in a multivariate setting.  Using the notation $S$ for scalars, $\mbf{V}$ for vectors, and $\msf{M}$ for matrices, what we require is $\msf{Y} = F (\msf{X})$, where $\msf{X}$ has size $[N_D, N_X]$ for $N_X$ points in $N_D$ dimensions and $\msf{Y}$ has size $[N_F, N_X]$ for $N_F$ integrands.  Vectorized expressions and binary singleton expansion operators $\oplus$ and $\otimes$ should be used when coding $F$.

The region of integration is specified by the vectors $\mbf{A}$ and $\mbf{B}$, each with $N_D$ elements, defining a hyper-rectangular volume such that \beql{eqn:R}
\mbf{R} = \int_\mbf{A}^\mbf{B} d\mbf{X}\, F (\mbf{X}) \pm \mbf{E}
\eeq is the vector of $N_F$ results we are after with estimated error $\mbf{E}$.  If necessary, the limits are swapped such that all $A_d < B_d$, accounting for any induced change of sign.  Optionally, a matrix of breakpoints $\msf{C}$ with size $[N_D, N_C]$ can be supplied by the user; if none is given the default $\msf{C}$ is determined to be the midpoint of the region.  The primary use of $\msf{C}$ is to inform AMGKQ of the locations of singular (or nearly singular) values of the integrand, but when doing complex line integrals $\msf{C}$ is used to define the path of the contour.

\begin{figure}
\centerline{\includegraphics[scale=.85]{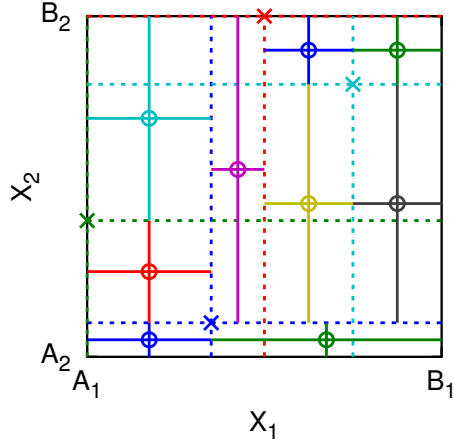}}
\caption{Initial subregions in 2 dimensions with four breakpoints, two of which are located on the outer boundary of the region.  Breakpoints are indicated by $\times$ and dotted lines, and subregions are indicated by $\bigcirc$ and solid lines.}
\label{fig:A}
\end{figure}

With $\msf{C}$ in hand, the initial subregions indexed by $s$ are defined in terms of their central location $\mbf{L}_s$ and half-width $\mbf{H}_s$ as follows.  That permutation of the $N_C$ locations in $\msf{C}$ which gives the shortest aggregate distance from $\mbf{A}$ through $\msf{C}$ to $\mbf{B}$ is selected.  Starting near $\mbf{A}$, each point $\mbf{C}_c$ in turn is used to subdivide the region in which it is found into $2^{N_D}$ partitions, discarding any null volumes.  Optionally, the user can request that $\msf{C}$ be taken in the order originally specified.  Points in $\msf{C}$ can be located on the outer boundary, $C_{d,c} = A_{d}$ or $C_{d,c} = B_{d}$, or on internal boundaries without fubaring the initial subdivision.  If all components in $\msf{C}$ are unique and not equal to any component of $\mbf{A}$ or $\mbf{B}$, one has $N_C ( 2^{N_D} - 1 ) + 1$ subregions when finished, which sets the lower limit on the requested maximum number of subregions AMGKQ is allowed to consider.  An example of the subdivision process in 2 dimensions is shown in Figure~\ref{fig:A}.

\subsection{Gauss-Kronrod Quadrature in Multiple Dimensions}

For any subregion labeled by $s$, we wish to compute the integral over the volume as efficiently as possible.  To do so, we store in memory the abscissa in normalized units (between -1 and 1) for every contributing location $\msf{X}_K$ of size $[N_D, N_K]$, as well as both their Gauss and Kronrod weights, $\mbf{W}_G$ and $\mbf{W}_K$ respectively.  The $(n_G, n_K)$ Gauss-Kronrod quadrature rule pair $(\mbf{w}_G, \mbf{w}_K)$ in one dimension $\mbf{x}_K$ can be of any order $n_K = 2 n_G + 1$, with tabulated values for those most commonly used \citep{Holoborodko-2011GK} and a double precision routine called for others \citep{Laurie-19971133,Gautschi-199421,Gautschi-2004}.  In multiple dimensions $N_D > 1$, one has $N_K = n_K^{N_D}$ weights in $\mbf{W}_K$ and $N_G = n_G^{N_D}$ weights in $\mbf{W}_G$.  Each element of $\mbf{W}_G$ and $\mbf{W}_K$ is the product of the weights in $\mbf{w}_G$ and $\mbf{w}_K$ respectively corresponding to the indexed location in $\msf{X}_K$, and $\msf{X}_G$ is the subset of $\msf{X}_K$ where every coordinate is of even parity.  The abscissa locations in physical units $\msf{X}_s$ can then be evaluated for the entire subregion by first calculating the locations along the central axes $\msf{X}_L = ( \mbf{H}_s \otimes \mbf{x}_K ) \oplus \mbf{L}_s$ and then constructing $\msf{X}_s$ from $\msf{X}_L$ by indexing.

The integrand is then evaluated at all locations $\msf{X}_s$ with a single call to the user supplied function, $\msf{Y}_s = F (\msf{X}_s)$.  The Kronrod estimate of the integral is calculated with a matrix multiplication and a scalar multiplication $\mbf{Q}_K = [ \msf{Y}_s (\msf{X}_K) \times \mbf{W}_K ] h_s$, where $h_s = \prod_d H_{d,s}$ is the volume factor, and similarly for the Gauss estimate $\mbf{Q}_K = [ \msf{Y}_s (\msf{X}_G) \times \mbf{W}_G ] h_s$.  The result for the subregion is given by the Kronrod estimate $\mbf{R}_s = \mbf{Q}_K$, and its variance is estimated as $\mbf{V}_s = ( \mbf{Q}_K - \mbf{Q}_G )^2$, taking the power along independent dimensions $f$.  The values $\mbf{R}_s$ and $\mbf{V}_s$ are stored in memory for accumulation.  The final result is the accumulation of all the subregion results $\mbf{R} = \sum_s \mbf{R}_s$, and its estimated error is the square root of the accumulated subregion variances $\mbf{E} = ( \sum_s \mbf{V}_s )^{1/2}$.  Note that ADAPT multiplies its estimate of $\mbf{E}$ by a factor of 3, while QUADGK accumulates the subregion errors $\mbf{V}_s^{1/2}$.

\subsection{Selecting the Subregion and Direction for Subdivision}

With each iteration of the main loop, that subregion among the $N_s$ present which has the single largest estimated variance across all integrands is selected for subdivision, while ADAPT selects the greatest error summed over integrands.  When evaluating the direction for subdivision for $N_D > 1$ according the magnitude of the fourth derivative of $F$, only that integrand with the largest variance is considered, while again ADAPT considers the sum over integrands.  Neither method is particularly well-suited when the orders of magnitude of the integrands are vastly different, but how best to make relative the selection process for simultaneous integrands is not clear.  When selecting the direction, AMGKQ focuses on the integrand that triggered the selection of the subregion.

The evaluation of the fourth derivative (in each dimension) is accomplished by using finite difference coefficients $\mbf{w}_4$.  Having evaluated $\msf{Y}_s$ as part of the Gauss-Kronrod quadrature, those values are used again for this purpose.  Since $\mbf{x}_K$ is not evenly spaced, the coefficients themselves must be calculated for the chosen order of quadrature rule \citep{Fornberg-1998685,LeVeque-2007}.  The abscissa locations $\msf{X}_d$ along the central axes are identified, and a matrix multiplication yields the result $\mbf{F}^{iv}_{d,s} = \msf{Y}_s (\msf{X}_d) \times \mbf{w}_4$ in normalized units.  That direction with the greatest fourth derivative in magnitude for the selected integrand is chosen for division by a factor of 2, such that each iteration contributes one additional subregion to the accumulation $N_s \leftarrow N_s + 1$.  The calculation of $\mbf{R}_s$, $\mbf{V}_s$, and $\mbf{F}^{iv}_{d,s}$ are implemented in the function $\gkintfunof{}$.

\subsection{Convergence, Subregion Culling, and Termination Criteria}

The user may request either or both an absolute tolerance $E_A$ and a relative tolerance $E_R$ for the convergence criterion.  When all components of $\mbf{E}$ are less than those of $\mbf{T}$, where $T_f$ is the greater of $E_A$ or $E_R \abs{R_f}$, the algorithm considers itself globally converged.  The estimated error $\mbf{E}$ is a measure of the precision of the result, which is not quite the same thing as accuracy.  To measure accuracy, one needs to know independently (analytically) what is the true value of the integral for comparison to its numeric approximation.  One hopes, of course, that the precision and accuracy will be of the same order of magnitude, but difficult integrands can lead one to a result that is precisely wrong.

The are two conditions under which a subregion may be culled from further consideration.  The first is when its estimated error is sufficiently small as to not affect (the current estimate of) the final result.  The second is when its half-width $\mbf{H}_s$ is approaching the limit of machine resolution in any dimension.  Technically, that condition is met when subdividing a half-width would result in a subregion whose outermost Kronrod abscissa in physical units is indistinguishable numerically from the subregion boundary; otherwise, the integrand might be evaluated at the location of a breakpoint.  When either condition is met, the contribution of the subregion to the accumulations is simply stored, and its location no longer considered.  During testing, the second condition arose only when estimating $\int_0^\infty dx \sin(x) / x = \pi / 2$, and the most accurate estimate of that integral was achieved by disabling the subregion culling entirely.  If no subregions remain after culling, the algorithm considers itself converged but does send a unique flag to the user.  The user also is warned if the second condition has been triggered.

There remain a few other conditions for which AMGKQ will terminate.  When the maximum number of subregions $N_S$ requested by the user have been considered, the algorithm will return a flag along with its last values for the result and estimated error.  Likewise, when a value of $\mrm{NaN}$ or $\pm \mrm{Inf}$ is encountered, the algorithm will terminate with specific flags.  These flags are meant to warn the user to inspect the result for accuracy.  The user also is warned if any of the estimated errors do not meet the requested precision upon termination.  The main loop of the algorithm can thus be stated as Algorithm~\ref{alg:mainloop}.

\begin{algorithm}[t]
\SetKwFunction{isempty}{isempty}
\SetKwFunction{gkint}{gkint}
\SetKwFunction{any}{any}
\SetKwFunction{all}{all}
\SetKw{KwOr}{or}
\SetAlgoNoLine
\KwIn{Initial subregions $\mbf{L}_s$, $\mbf{H}_s$, and function $F(x)$.}
\KwOut{Final estimates $\mbf{R}$, $\mbf{E}$, subregions evaluated $N_s$, and flag $I$.}
\lForAll{s}{$[ \mbf{R}_s, \mbf{V}_s, \msf{F}^{iv}_s ]$ = \gkint{$\mbf{L}_s, \mbf{H}_s, F(x)$}}\;
also compute culling tolerances for all $\mbf{H}_s$\;
$\mbf{R}$ = $\sum_s \mbf{R}_s$; $\mbf{V}$ = $\sum_s \mbf{V}_s$\;
$first$ = $TRUE$\;
\While{$N_s \leq N_S$}{
		\lIf{\any{$\mbf{R}_s$ == $\mrm{NaN}$ \KwOr $\mrm{Inf}$}}{break}\;
		evaluate tolerance vector $\mbf{T}$\;
		\lIf{\all{$\mbf{V} < \mbf{T}^2$}}{break}\;
		find indices $\{s'\} \subset \{s\}$ for culling\;
		\eIf{first}{
			$\mbf{R}'$ = $\sum_{s'} \mbf{R}_{s'}$; $\mbf{V}'$ = $\sum_{s'} \mbf{V}_{s'}$\;
			$first$ = $FALSE$\;
		}{
			$\mbf{R}$ = $\mbf{R}'$ + $\sum_s \mbf{R}_s$; $\mbf{V}$ = $\mbf{V}'$ + $\sum_s \mbf{V}_s$\; 
			$\mbf{R}'$ += $\sum_{s'} \mbf{R}_{s'}$; $\mbf{V}'$ += $\sum_{s'} \mbf{V}_{s'}$\;
		}
		remove indices $\{s'\}$ from $\{s\}$\;
		\lIf{\isempty{$\{s\}$}}{break}\;
		find $s'' \in \{s\}$ and integrand $f$ with greatest error\;
		select direction $d$ for subdivision\;
		$N_s$ += 1; $s'''$ = $s_\mrm{max}$ + 1\;
		$H_{d,s''}$ = $H_{d,s''}$ / 2\;
		$\mbf{L}_{s'''}$ = $\mbf{L}_{s''}$; $\mbf{H}_{s'''}$ = $\mbf{H}_{s''}$\;
		$L_{d,s''}$ = $L_{d,s''}$ - $H_{d,s''}$; $L_{d,s'''}$ = $L_{d,s'''}$ + $H_{d,s'''}$\;
		$[ \mbf{R}_{s''}, \mbf{V}_{s''}, \msf{F}^{iv}_{s''} ]$ = \gkint{$\mbf{L}_{s''}, \mbf{H}_{s''}, F(x)$}\;
		$[ \mbf{R}_{s'''}, \mbf{V}_{s'''}, \msf{F}^{iv}_{s'''} ]$ = \gkint{$\mbf{L}_{s'''}, \mbf{H}_{s'''}, F(x)$}\;
		also compute culling tolerances for $\mbf{H}_{s''}$ and $\mbf{H}_{s'''}$\;
      }
      account for sign of $\mbf{R}$ and take square root $\mbf{E}$ = $\mbf{V}^{1/2}$\;
      set flag $I$ and express warnings\;
\caption{Main Loop}
\label{alg:mainloop}
\end{algorithm}

\subsection{Contour Integrals}

Complex line (contour) integrals can be accomplished using the same machinery with no changes beyond some additional $\absfunof{}$ functions that appear in the evaluation of the half-width tolerances.  These integrals are restricted to $N_D = 1$, which is understood to represent a single complex plane, and finite values for all components of $\mbf{A}$, $\mbf{B}$, and $\msf{C}$.  The points $\mbf{A}$ and $\mbf{B}$ are the starting and ending points of integration (which usually will be equal but are not required to be so), and the points in $\msf{C}$ determine the path of integration in a piecewise continuous linear fashion; no reordering of $\msf{C}$ is done in this case.  Everything else proceeds the same as for the case of real integrals.

\section{Improper Integrals and Variable Transformations}

The algorithm AMGKQ can handle integrals that are improper, either because the integrand diverges at the boundary of the region or the domain of integration is itself unbounded.  Singularities within the domain should be avoided by use of breakpoints.  As long as the integrand is sufficiently well behaved, the result will be an accurate approximation to the value of the integral.

\subsection{Edge Singularities}

\begin{figure}
\centerline{\includegraphics[scale=.8]{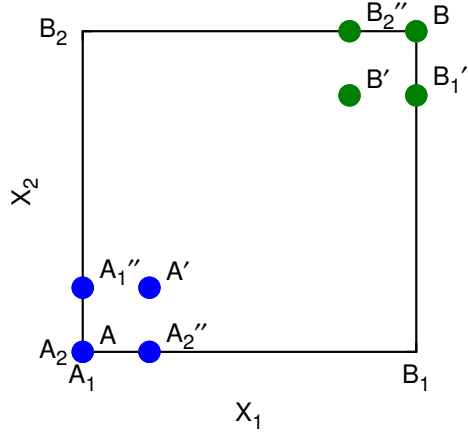}}
\caption{Schematic for $N_D = 2$ of how the dimensions corresponding to edge singularities are determined.}
\label{fig:B}
\end{figure}

If a singular integrand is detected at either or both $\mbf{A}$ and $\mbf{B}$, a variable transformation of the form $\int dx\, F(x) = \int dy\, F(x_y)\, dx_y/dy$ is employed to weaken the singularity; no attempt is made to weaken singularities at points in $\msf{C}$.  To find which dimensions are causing the singularity, a heuristic algorithm is employed.  The idea is to define a point $\mbf{A}'$ which is ``near $\mbf{A}$'', and similarly for $\mbf{B}'$.  In case of infinite limits, one must check that ``near $\mbf{A}$'' is also ``far from $\mbf{B}$'' and act accordingly.  Then, for each direction $d$, one replaces $A_d'$ with $A_d$ to form $\mbf{A}_d''$ and inspects $\mbf{Y}_{d,\mbf{A}}'' = F (\mbf{A}_d'')$, and similarly for $\mbf{Y}_{d,\mbf{B}}''$.  Any integrand which is not finite triggers the need for a variable transformation in that dimension at either or both endpoints as necessary.  If the endpoints are themselves infinite in those dimensions, AMGKQ complains that the integral is divergent and throws an error.  If not, the algorithm proceeds to effect the variable transformations.  A schematic depiction for $N_D = 2$ is shown in Figure~\ref{fig:B}.

\subsubsection{Both A and B}

This case is the most complicated thus will be considered first.  Let $\{d'\} \subset \{d\}$ be those dimensions for which singular integrands are detected at both $A_{d'}$ and $B_{d'}$.  The user has the option of selection either a trigonometric or a rational function for the transformation (in all $d'$, not independently).  To accomplish the transformation, one needs to know $x_y \equiv x(y)$ and $dx_y/dy$, as well as $y_x \equiv y(x)$ to find the new limits $A_{d'}$ and $B_{d'}$ and the breakpoints $\msf{C}_{d'}$ in the new geometry.  For the trigonometric transformation, those functions are \bes
x_{d'} (y_{d'}) &=& [ 1 - \cos (y_{d'}) ] ( B_{d'} - A_{d'} ) / 2 + A_{d'} \;, \\
dx_{d'} / dy_{d'} &=& \sin (y_{d'}) ( B_{d'} - A_{d'} ) / 2 \;, \\
y_{d'} (x_{d'}) &=& 2 \arctan \{ [x_{d'}' / (1 - x_{d'}')]^{1/2} \} \; ,
\ees where $x_{d'}' = ( x_{d'} - A_{d'} ) / ( B_{d'} - A_{d'} )$.  For the rational transformation we have \bes
x_{d'} (y_{d'}) &=& y_{d'} ( 3 - y_{d'}^2 ) ( B_{d'} - A_{d'} ) / 4 + ( B_{d'} + A_{d'} ) / 2 \;, \\
dx_{d'} / dy_{d'} &=& 3 ( 1 - y_{d'}^2 ) ( B_{d'} - A_{d'} ) / 4 \;, \\
y_{d'} (x_{d'}) &=& [ \sqrt{-3} ( 1 - x_{d'}''^2 ) - ( 1 + x_{d'}''^2 ) ] / 2 x_{d'}'' \; ,
\ees where $x_{d'}'' = \{ [ ( x_{d'}'^2 - 4 )^{1/2} + x_{d'}' ] / 2 \}^{1/3}$ and $x_{d'}' = [ 2 ( B_{d'} + A_{d'} ) - 4 x_{d'} ] / ( B_{d'} - A_{d'} )$; the imaginary part of $y_{d'}$ should be 0 to machine precision and can be discarded.  For either transformation, one can construct the transformed integrand efficiently using binary singleton expansion such that $F (\msf{X}') = \big[ \prod_{d'} ( dx_{d'} / dy_{d'} )_{\msf{X}'} \big] \otimes F (\msf{X}_{\msf{X}'})$.

\subsubsection{Just A or Just B}  For these two cases only a rational transformation is available to the user. When $\{d''\}$ is the set of dimensions for which singular integrands are detected only at $A_{d''}$, the transformation functions are \bes
x_{d''} (y_{d''}) &=& A_{d''} + y_{d''}^2 \;, \\
dx_{d''} / dy_{d''} &=& 2 y_{d''} \;, \\
y_{d''} (x_{d''}) &=& ( x_{d''} - A_{d''} )^{1/2} \;,
\ees and when $\{d'''\}$ is the set of dimensions for which singular integrands are detected only at $B_{d'''}$, the transformation functions are \bes
x_{d'''} (y_{d'''}) &=& B_{d'''} - y_{d'''}^2 \;, \\
dx_{d'''} / dy_{d'''} &=& - 2 y_{d'''} \;, \\
y_{d'''} (x_{d'''}) &=& - ( B_{d'''} - x_{d'''} )^{1/2} \;.
\ees  All three cases are processed sequentially, which can lead to a final function of the form \beq
F (\msf{X}''') = \left[ \prod_{d'''} ( dx_{d'''} / dy_{d'''} )_{\msf{X}'''} \right] \otimes \left[ \prod_{d''} ( dx_{d''} / dy_{d''} )_{\msf{X}''} \right] \otimes \left[ \prod_{d'} ( dx_{d'} / dy_{d'} )_{\msf{X}'} \right] \otimes F (\msf{X}_{\msf{X}'''}) \;,
\eeq that gets passed to the main loop performing the actual quadrature, after accounting for any infinite limits that may be present.

\subsection{Infinite Limits}

When infinite limits appear in either $\mbf{A}$ or $\mbf{B}$, a variable transformation is employed to map the manifold to a finite domain.  The user has the option of selecting either a trigonometric or rational function for the transformation.  Let $\{d''''\}$ be the set of dimensions which have at least one infinite limit.  For the trigonometric transformation, the required functions are \bes
x_{d''''} (y_{d''''}) &=& \tan ( y_{d''''} ) \;, \\
dx_{d''''} / dy_{d''''} &=& \sec^2 ( y_{d''''} ) \;, \\
y_{d''''} (x_{d''''}) &=& \arctan ( x_{d''''} ) \;,
\ees and the required functions for the rational transformation are \bes
x_{d''''} (y_{d''''}) &=& y_{d''''} / ( 1 - y_{d''''}^2 ) \;, \\
dx_{d''''} / dy_{d''''} &=& ( 1 + y_{d''''}^2 ) / ( 1 - y_{d''''}^2 )^2 \;, \\
y_{d''''} (x_{d''''}) &=& 2 x_{d''''} / [ 1 + ( 1 + 4 x_{d''''}^2 )^{1/2} ] \;.
\ees  The infinite limit transformation $F (\msf{X}'''') = \big[ \prod_{d''''} ( dx_{d''''} / dy_{d''''} )_{\msf{X}''''} \big] \otimes F (\msf{X}'''_{\msf{X}''''})$ is applied after any arising from edge singularities.  If the default breakpoint $\mbf{C} = (\mbf{A} + \mbf{B})/2$ is selected because none were specified by the user, one must account for infinite limits by taking $C_{d''''} = (A_{d''''} + B_{d''''})/2$ in the new coordinates.

\section{Accuracy Testing}

To test the accuracy of any numeric integration algorithm, one must assemble a collection of integrals whose values are known exactly.  A well known collection is provided by John Burkardt \citeyear{Burkardt-2009,Burkardt-2011}, a subset of which will be used here, sometimes modified for convenience.  The analytic forms of these integrals have been included in the documentation provided with the code, as have example scripts that generate the results shown here.  Some typos in the exact values quoted in the first library have been corrected, which are now evaluated in terms of their closed form solution.

\subsection{Burkardt Tests}

A set of 31 functions is selected for testing in one dimension as displayed in Table~\ref{tab:A}.  The integrand is passed to AMGKQ in the form of an anonymous function handle.  Default parameters of $E_A = \sqrt{\epsof{1}} \approx$ 1.5e-8, $E_R = 0$, and $N_S = 2^{N_D} \times 100$ control the algorithm, where $\epsof{x}$ is the floating point resolution of value $x$.  The limits for these functions are all finite.  The number of subregions evaluated $N_s$ is displayed, as is the output flag: 2 means globally converged, 1 means locally converged, 0 means $N_S$ is reached, and $<$0 means $\mrm{Inf}$ or $\mrm{NaN}$ has been encountered.  The estimated error is under the heading ERR, and the actual accuracy is under ACC.

\begin{table}[]\footnotesize
\caption{Burkardt Tests for $N_D = 1$\label{tab:A}}
\centering
\begin{tabular}{l|ccc|cc|cc}
\hline
No. & $ F(X) $ & $A$ & $B$ & $N_s$ & flag & ERR & ACC \\\hline 
1 & $ \exp (X) $ & 0.0 & 1.0 & 2 & 2 & 1.1e-16 & 2.2e-16 \\ 
2 & $ 1 / (1 + X^4) $ & 0.0 & 1.0 & 2 & 2 & 7.7e-13 & 0.0e+00 \\ 
3 & $ 1 / (1 + \exp (X)) $ & 0.0 & 1.0 & 2 & 2 & 0.0e+00 & 0.0e+00 \\ 
4 & $ X / (\exp (X) - 1) $ & 0.0 & 1.0 & 2 & 2 & 3.9e-16 & 8.9e-16 \\ 
5 & $ X / (\exp (X) + 1) $ & 0.0 & 1.0 & 2 & 2 & 0.0e+00 & 2.8e-17 \\ 
6 & $ 0.92 \cosh (X) - \cos (X) $ & -1.0 & 1.0 & 2 & 2 & 3.9e-17 & 2.2e-16 \\ 
7 & $ \exp (X) \cos (X) $ & 0.0 & 3.1 & 2 & 2 & 5.1e-14 & 0.0e+00 \\ 
8 & $ 1 / (1 + X^2 + X^4) $ & -1.0 & 1.0 & 2 & 2 & 6.6e-09 & 6.7e-16 \\ 
9 & $ 50 / \pi / (2500 X^2 + 1) $ & 0.0 & 1.0 & 8 & 2 & 9.7e-12 & 5.6e-17 \\ 
10 & $ \sqrtfun (X) $ & 0.0 & 1.0 & 12 & 2 & 7.1e-09 & 4.1e-10 \\ 
11 & $ \sqrtfun (50) \exp (-50 \pi X^2) $ & 0.0 & 10.0 & 8 & 2 & 2.3e-09 & 1.1e-16 \\ 
12 & $ 25 \exp (-25 X) $ & 0.0 & 10.0 & 8 & 2 & 2.0e-11 & 0.0e+00 \\ 
13 & $ 1 / \sqrtfun (X) $ & 0.0 & 1.0 & 2 & 2 & 2.2e-16 & 0.0e+00 \\ 
14 & $ \log (X) $ & 0.0 & 1.0 & 10 & 2 & 9.7e-09 & -1.8e-10 \\ 
15 & $ \sqrtfun (\absfun (X + 0.5)) $ & -1.0 & 1.0 & 22 & 2 & 1.0e-08 & 8.2e-10 \\ 
16 & $ \log (\absfun (X - 0.7)) $ & 0.0 & 1.0 & 28 & 2 & 1.3e-08 & -3.4e-09 \\ 
17 & $ 2 / (2 + \sin (10 \pi X)) $ & 0.0 & 1.0 & 17 & 2 & 1.3e-08 & -2.6e-14 \\ 
18 & $ (\sin (50 \pi X))^2 $ & 0.0 & 1.0 & 5 & 2 & 3.4e-16 & -1.1e-16 \\ 
19 & $ \exp (\cos (X)) $ & 0.0 & 6.3 & 5 & 2 & 1.4e-10 & 8.9e-16 \\ 
20 & $ 1 / (X^{1/2} + X^{1/3}) $ & 0.0 & 1.0 & 15 & 2 & 8.6e-09 & 7.3e-10 \\ 
21 & $ \exp (-X) \sin (50 X) $ & 0.0 & 6.3 & 52 & 2 & 1.3e-08 & -4.5e-17 \\ 
22 & $ (X <= \exp (1) - 2) / (X + 2) $ & 0.0 & 1.0 & 22 & 2 & 1.1e-08 & 3.1e-09 \\ 
23 & $ 1 / (1 + X^2) $ & -4.0 & 4.0 & 7 & 2 & 9.4e-10 & -4.4e-16 \\ 
24 & $ \sqrtfun (-\log (X)) $ & 0.0 & 1.0 & 18 & 2 & 1.0e-08 & 6.0e-10 \\ 
25 & $ \prod_{k = 0}^3 (10 x - 1 - k / 10) $ & 0.0 & 1.0 & 2 & 2 & 0.0e+00 & 2.3e-13 \\ 
26 & $ \log (X) \sqrtfun (X) $ & 0.0 & 1.0 & 14 & 2 & 1.2e-08 & -7.8e-10 \\ 
27 & $ \log (X) / \sqrtfun (X) $ & 0.0 & 1.0 & 24 & 2 & 1.3e-08 & 2.3e-09 \\ 
28 & $ (0.3 <= X) $ & 0.0 & 1.0 & 24 & 2 & 9.3e-09 & 8.2e-09 \\ 
29 & $ \sum_{k = 1}^3 (\sech (10^k (X - k / 5)))^{2 k} $ & 0.0 & 1.0 & 11 & 2 & 3.9e-09 & -1.1e-03 \\ 
30 & $ \sum_{k = 1}^{40} \cos (7^k X \pi / 2) / 2^k $ & 0.0 & 1.0 & 200 & 0 & 2.1e-04 & 1.1e-03 \\ 
31 & $ (1 / X) \sin (1 / X) $ & 0.0 & 1.0 & 200 & 0 & 4.3e-01 & 7.2e-01 \\ 
\hline 
29 & $ \sum_{k = 1}^3 (\sech (10^k (X - k / 5)))^{2 k} $ & 0.0 & 1.0 & 57 & 1 & 4.5e-17 & 0.0e+00 \\ 
30 & $ \sum_{k = 1}^{40} \cos (7^k X \pi / 2) / 2^k $ & 0.0 & 1.0 & 1000 & 0 & 8.3e-05 & -1.5e-05 \\ 
31 & $ (1 / X) \sin (1 / X) $ & 0.0 & 1.0 & 1000 & 0 & 1.4e-01 & -2.2e-01 \\ 

\hline
\end{tabular}
\end{table}%

\begin{figure}[b]
\centerline{\includegraphics[scale=.8]{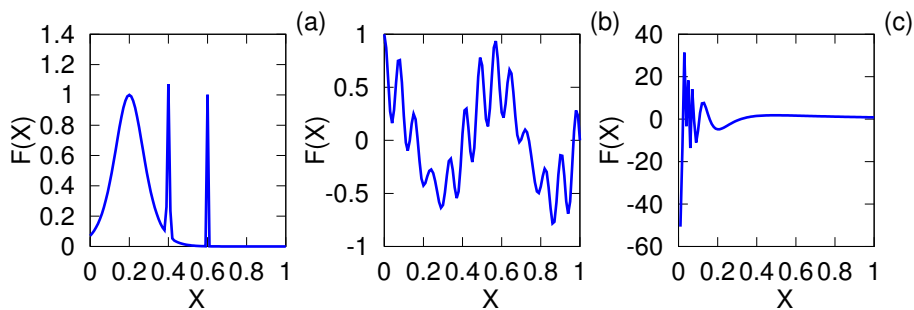}}
\caption{Difficult integrands in one dimension.}
\label{fig:C}
\end{figure}

For the vast majority of the selected functions, AMGKQ performs brilliantly.  Only for the three functions numbered 29, 30, and 31 is ACC above the requested precision.  The first is an example of a result that is precisely wrong, while the other two have an estimated error on par with their accuracy.  The integration is repeated for these functions, which are displayed in Figure~\ref{fig:C}, with parameters $E_A = 0$ and $N_S = 1000$, and results are appended to the bottom of the table; what makes these integrands difficult are the sharp peaks in panel (a), the low-amplitude, high-frequency content in panel (b), and the wild oscillations in panel (c).  Function number 29 is evaluated accurately when forced to converge locally, while function number 30 improves with more iterations.  Function number 31 can be related to the sine integral which will be discussed later.

In two dimensions \citet{Burkardt-2011} does not give as many functions to investigate.  These integrals all have limits which are the same for both directions, so only one value will be displayed for $\mbf{A}$ and $\mbf{B}$.  The results of the accuracy tests following the same procedure as above are shown in Table~\ref{tab:B}.  Note that the two difficult integrands which were repeated both contain $\absfunof{}$ as part of the operation; the discontinuity in the first derivative apparently makes high accuracy hard to achieve.  Also note that the Gaussian function, which is representative of what is encountered in Bayesian data analysis, converges quickly compared to the others.

\begin{table}[]\footnotesize
\caption{Burkardt Tests for $N_D = 2$\label{tab:B}}
\centering
\begin{tabular}{l|ccc|cc|cc}
\hline
No. & $ F(X,Y) $ & $A$ & $B$ & $N_s$ & flag & ERR & ACC \\\hline 
1 & $ 1 / (1 - x y) $ & 0.0 & 1.0 & 43 & 2 & 1.2e-08 & -2.0e-09 \\ 
2 & $ 1 / \sqrtfun (1 - x^2 y^2) $ & -1.0 & 1.0 & 92 & 2 & 1.4e-08 & -1.4e-09 \\ 
3 & $ 1 / \sqrtfun (2 - x - y) $ & -1.0 & 1.0 & 27 & 2 & 7.1e-09 & -3.7e-10 \\ 
4 & $ 1 / \sqrtfun (3 - x - 2 y) $ & -1.0 & 1.0 & 26 & 2 & 1.0e-08 & -5.3e-10 \\ 
5 & $ \sqrtfun (x y) $ & 0.0 & 1.0 & 67 & 2 & 1.3e-08 & 2.5e-09 \\ 
6 & $ \exp (-((x - 4)^2 + (y - 1)^2)) $ & 0.0 & 5.0 & 10 & 2 & 6.0e-09 & 4.4e-16 \\ 
7 & $ \absfun (x^2 + y^2 - 0.25) $ & -1.0 & 1.0 & 379 & 2 & 1.5e-08 & 4.4e-08 \\ 
8 & $ \sqrtfun (\absfun (x - y)) $ & 0.0 & 1.0 & 400 & 0 & 3.2e-07 & -1.0e-06 \\\hline
7 & $ \absfun (x^2 + y^2 - 0.25) $ & -1.0 & 1.0 & 1000 & 0 & 7.8e-10 & 6.9e-09 \\ 
8 & $ \sqrtfun (\absfun (x - y)) $ & 0.0 & 1.0 & 1000 & 0 & 5.0e-08 & -2.3e-07 \\ 

\hline
\end{tabular}
\end{table}%

\subsection{Further Tests}

\begin{table}[]\footnotesize
\caption{Further Accuracy Tests\label{tab:C}}
\centering
\begin{tabular}{l|ccc|cc|cc}
\hline
No. & $ F(\mbf{X}) $ & $\mbf{A}$ & $\mbf{B}$ & $N_s$ & flag & ERR & ACC \\\hline 
1 & $ \begin{array}{c} 1/(1 + z^2)^2 \\  \exp(i z)/(1 + z^2)\end{array} $ & $ \begin{tabular}{c} -1.0 \end{tabular} $ & $ \begin{tabular}{c} -1.0 \end{tabular} $ & 13 & 2 & $ \begin{tabular}{c} 5.0e-09-2.0e-17i \\ 1.4e-09-3.3e-17i \end{tabular} $ & $ \begin{tabular}{c} -2.2e-16+2.8e-17i \\ 0.0e+00+4.2e-17i \end{tabular} $ \\ \hline
2 & $ 1/\sqrtfun (\absfun (x)) $ & $ \begin{tabular}{c} 0.0 \end{tabular} $ & $ \begin{tabular}{c} 10.0 \end{tabular} $ & 2 & 2 & $ \begin{tabular}{c} 2.2e-16 \end{tabular} $ & $ \begin{tabular}{c} 0.0e+00 \end{tabular} $ \\ 
3 & $ 1/\sqrtfun (\absfun (x)) $ & $ \begin{tabular}{c} -10.0 \end{tabular} $ & $ \begin{tabular}{c} 10.0 \end{tabular} $ & 101 & 2 & $ \begin{tabular}{c} 1.3e-08 \end{tabular} $ & $ \begin{tabular}{c} -1.2e-08 \end{tabular} $ \\ 
4 & $ 1/(\sqrtfun (x) (1 + x)) $ & $ \begin{tabular}{c} 0.0 \end{tabular} $ & $ \begin{tabular}{c} $\infty$ \end{tabular} $ & 2 & 2 & $ \begin{tabular}{c} 2.2e-16 \end{tabular} $ & $ \begin{tabular}{c} 0.0e+00 \end{tabular} $ \\ 
5 & $ \log (x)/(1 - x^2) $ & $ \begin{tabular}{c} 0.0 \end{tabular} $ & $ \begin{tabular}{c} 1.0 \end{tabular} $ & 10 & 2 & $ \begin{tabular}{c} 9.7e-09 \end{tabular} $ & $ \begin{tabular}{c} -1.8e-10 \end{tabular} $ \\ 
6 & $ \exp (-x) x/(1 - \exp (-2 x)) $ & $ \begin{tabular}{c} 0.0 \end{tabular} $ & $ \begin{tabular}{c} $\infty$ \end{tabular} $ & 6 & 2 & $ \begin{tabular}{c} 1.4e-08 \end{tabular} $ & $ \begin{tabular}{c} 2.7e-13 \end{tabular} $ \\ \hline
7 & $ \begin{array}{c} \exp(-x) x \\  \exp(-x) x^2 \\  \exp(-x) x^3 \\  \exp(-x) x^4 \\  \exp(-x) x^5\end{array} $ & $ \begin{tabular}{c} 0.0 \end{tabular} $ & $ \begin{tabular}{c} $\infty$ \end{tabular} $ & 12 & 2 & $ \begin{tabular}{c} 3.4e-13 \\ 3.3e-13 \\ 8.3e-11 \\ 1.5e-09 \\ 6.8e-09 \end{tabular} $ & $ \begin{tabular}{c} 0.0e+00 \\ 4.4e-16 \\ 3.6e-15 \\ 2.1e-14 \\ 1.6e-13 \end{tabular} $ \\ \hline
8 & $ \exp (-x^2) $ & $ \begin{tabular}{c} -$\infty$ \end{tabular} $ & $ \begin{tabular}{c} $\infty$ \end{tabular} $ & 9 & 2 & $ \begin{tabular}{c} 1.8e-09 \end{tabular} $ & $ \begin{tabular}{c} 2.2e-16 \end{tabular} $ \\ 
9 & $ \exp (-x^2) cos (x) $ & $ \begin{tabular}{c} 0.0 \end{tabular} $ & $ \begin{tabular}{c} $\infty$ \end{tabular} $ & 5 & 2 & $ \begin{tabular}{c} 4.1e-09 \end{tabular} $ & $ \begin{tabular}{c} 1.1e-16 \end{tabular} $ \\ 
10 & $ \exp (-x^2)(1 + x^2)^{-1} $ & $ \begin{tabular}{c} 0.0 \end{tabular} $ & $ \begin{tabular}{c} 1.0 \end{tabular} $ & 2 & 2 & $ \begin{tabular}{c} 2.2e-13 \end{tabular} $ & $ \begin{tabular}{c} 2.2e-16 \end{tabular} $ \\ \hline
11 & $ \exp (-x_1^2 / 2)(1 + x_2^2)^{-1} $ & $ \begin{tabular}{c} -$\infty$ \\ -$\infty$ \end{tabular} $ & $ \begin{tabular}{c} $\infty$ \\ $\infty$ \end{tabular} $ & 17 & 2 & $ \begin{tabular}{c} 6.0e-09 \end{tabular} $ & $ \begin{tabular}{c} 4.3e-14 \end{tabular} $ \\ \hline
12 & $ \exp (-x_1^2 / 2)(1 + x_2^2)^{-1} $ & $ \begin{tabular}{c} -10.0 \\ -10.0 \end{tabular} $ & $ \begin{tabular}{c} 10.0 \\ 10.0 \end{tabular} $ & 76 & 2 & $ \begin{tabular}{c} 1.3e-08 \end{tabular} $ & $ \begin{tabular}{c} -2.1e-11 \end{tabular} $ \\ \hline
13 & $ \begin{array}{c} \exp(-x_1^2 / 2) \\  (1 + x_2^2)^{-1}\end{array} $ & $ \begin{tabular}{c} -10.0 \\ -10.0 \end{tabular} $ & $ \begin{tabular}{c} 10.0 \\ 10.0 \end{tabular} $ & 56 & 2 & $ \begin{tabular}{c} 1.0e-08 \\ 1.5e-08 \end{tabular} $ & $ \begin{tabular}{c} 5.7e-14 \\ -1.4e-14 \end{tabular} $ \\ \hline
14 & $ \begin{array}{l} \exp (-x_1^2 / 2)(1 + x_2^2)^{-1} \\ \quad \times x_3^{10} (1 - x_3)^{10} \end{array} $ & $ \begin{tabular}{c} -10.0 \\ -10.0 \\ 0.0 \end{tabular} $ & $ \begin{tabular}{c} 10.0 \\ 10.0 \\ 1.0 \end{tabular} $ & 8 & 2 & $ \begin{tabular}{c} 9.9e-10 \end{tabular} $ & $ \begin{tabular}{c} -9.0e-12 \end{tabular} $ \\ \hline
15 & $ \begin{array}{c} \exp(-x_1^2 / 2) \\  (1 + x_2^2)^{-1} \\  x_3^{10} (1 - x_3)^{10}\end{array} $ & $ \begin{tabular}{c} -10.0 \\ -10.0 \\ 0.0 \end{tabular} $ & $ \begin{tabular}{c} 10.0 \\ 10.0 \\ 1.0 \end{tabular} $ & 104 & 2 & $ \begin{tabular}{c} 7.4e-09 \\ 1.5e-08 \\ 2.4e-11 \end{tabular} $ & $ \begin{tabular}{c} 1.1e-13 \\ -3.6e-14 \\ -4.3e-19 \end{tabular} $ \\ \hline
16 & $ x^{-1/2} (1 - x)^{-1/2} $ & $ \begin{tabular}{c} 0.0 \end{tabular} $ & $ \begin{tabular}{c} 1.0 \end{tabular} $ & 2 & 2 & $ \begin{tabular}{c} 3.5e-14 \end{tabular} $ & $ \begin{tabular}{c} 9.8e-15 \end{tabular} $ \\ 
17 & $ x^{-2/3} (1 - x)^{-2/3} $ & $ \begin{tabular}{c} 0.0 \end{tabular} $ & $ \begin{tabular}{c} 1.0 \end{tabular} $ & 40 & -1 & $ \begin{tabular}{c} 2.9e-06 \end{tabular} $ & $ \begin{tabular}{c} -1.4e-05 \end{tabular} $ \\ 
18 & $ x^{-3/4} (1 - x)^{-3/4} $ & $ \begin{tabular}{c} 0.0 \end{tabular} $ & $ \begin{tabular}{c} 1.0 \end{tabular} $ & 40 & -1 & $ \begin{tabular}{c} 1.4e-04 \end{tabular} $ & $ \begin{tabular}{c} -4.8e-04 \end{tabular} $ \\ 
19 & $ (\sin (x)/x)^2 $ & $ \begin{tabular}{c} 0.0 \end{tabular} $ & $ \begin{tabular}{c} $\infty$ \end{tabular} $ & 200 & 0 & $ \begin{tabular}{c} 4.5e-06 \end{tabular} $ & $ \begin{tabular}{c} 5.9e-06 \end{tabular} $ \\ 
20 & $ (\sin (x)/x)^3 $ & $ \begin{tabular}{c} 0.0 \end{tabular} $ & $ \begin{tabular}{c} $\infty$ \end{tabular} $ & 146 & 2 & $ \begin{tabular}{c} 1.5e-08 \end{tabular} $ & $ \begin{tabular}{c} -7.6e-08 \end{tabular} $ \\ 
21 & $ (\sin (x)/x)^4 $ & $ \begin{tabular}{c} 0.0 \end{tabular} $ & $ \begin{tabular}{c} $\infty$ \end{tabular} $ & 35 & 2 & $ \begin{tabular}{c} 1.5e-08 \end{tabular} $ & $ \begin{tabular}{c} -2.0e-09 \end{tabular} $ \\ \hline
22 & $ \begin{array}{c} (\sum_d x_d^2 < 1) \\  (\sum_d x_d^2 > 1)\end{array} $ & $ \begin{tabular}{c} -1.0 \\ -1.0 \end{tabular} $ & $ \begin{tabular}{c} 1.0 \\ 1.0 \end{tabular} $ & 400 & 0 & $ \begin{tabular}{c} 7.0e-05 \\ 7.0e-05 \end{tabular} $ & $ \begin{tabular}{c} 6.9e-05 \\ -6.9e-05 \end{tabular} $ \\ 

\hline
\end{tabular}
\end{table}%

We now turn our attention to a set of functions selected from those used during development, displayed in Table~\ref{tab:C}.  It includes examples of simultaneous integrands and multiple dimensions; to guide the eye, horizontal rules distinguish those from single integrands in one dimension.  It also includes examples of improper integrals of either type, as well as an example of contour integration.  For function numbered 1, the breakpoints are $\msf{C} = [1, 2i]$ and the exact values are $\pi/2$ and $\pi/e$.  Default parameters were passed to AMGKQ for all these tests.  The effect of not ameliorating internal singularities can be seen by comparing numbers 2 and 3.

Difficult integrands have been collected at the bottom of the table.  Edge singularities stronger than $x^{-1/2}$, functions 17 and 18, are seen to terminate after encountering a value of $\mrm{Inf}$, returning a result that is not wildly off the mark.  Powers of the sinc function are also seen to converge, though number 19 requires more iterations than it was allowed.  The strong discontinuity of number 22 is not well modeled by the Gauss-Kronrod interpolating polynomial, such that the desired precision is hard to reach.

The numeric approximation of the integral of the sinc function $f(x) = \sin(x) / x$ over the semi-infinite domain is notoriously difficult.  We have encountered one of its forms before among the difficult integrands of the Burkardt tests.  If we ask QUADGK to approximate $\int_0^\infty dx \sin(x) / x = \pi / 2 \approx 1.5708$, it returns $R =  5.7135$ and $E = 7.14403$.  If we call AMGKQ with $N_S = 1000$ and disable subregion culling, we get a value of $R =  1.5570$, which has a relative accuracy of less than 1\%.  Speaking of relative accuracy, if we evaluate $\int_{10}^{15} dx \sin (3 x) \cosh (x) \sinh (x) \approx$ 2.6e+10, with $E_A = 0$ and $E_R =$ 1e-14, we find that the relative accuracy is indeed less than the requested relative precision.

\section{Performance Testing}

Having considered the advice given by \citet{Johnson-2002215}, let us embark on some performance testing.  We will investigate the accuracy and running time in two and three dimensions of four functions, three of which are localized and one of which is oscillatory.  In the order considered, the functions are a product over dimension of normal distributions $F(\mbf{x}) = \prod_d \exp (-x_d^2)$, a product of Cauchy distributions $F(\mbf{x}) = \prod_d ( 1 + x_d^2 )^{-1}$, a product of beta distributions $F(\mbf{x}) = \prod_d \exp (2 x_d) / [1 + \exp (x_d) ]^4$, and a product of squared sinusoids $F(\mbf{x}) = \prod_d \sin^2 (x_d) \cos^2 (x_d)$.  The integration region begins as a square (or hyper-square) with sides of length 2 units, whose center is offset from the origin by up to half a unit in any direction, and is scaled by integer factors of $k$ for successive runs.  The normalization of each integral is set to unity for each $k$ for consistency of comparison. 

For two dimensional integrals $N_D = 2$, Octave provides DBLQUAD, which calls recursively a chosen one dimensional quadrature routine such that vectorization of the integrand is only necessary for the first direction.  For our purposes, we select QUADGK and QUADCC as our integration routines; QUADCC implements Clenshaw-Curtis quadrature rules.  The adaptive Lobatto routine QUADL was abandoned for failing to terminate within a reasonable time when $k$ becomes large, as was QUADV using an adaptive Simpson's rule.  We also include ADAPT in our comparison to be complete, modified slightly to ignore the number of function evaluations and instead track $N_s$.  All quadrature routines are called with their default parameters for this test, except that they have $E_A =$ 1.0e-8 in common.  The testing environment is Octave 3.8.1 using the ATLAS BLAS library running on a Pentium 4 CPU at 3.0 GHz with 3GB of RAM.  The results are averaged over 3 trials to reduce their stochastic fluctuation.

\begin{figure}
\centerline{\includegraphics[scale=.8]{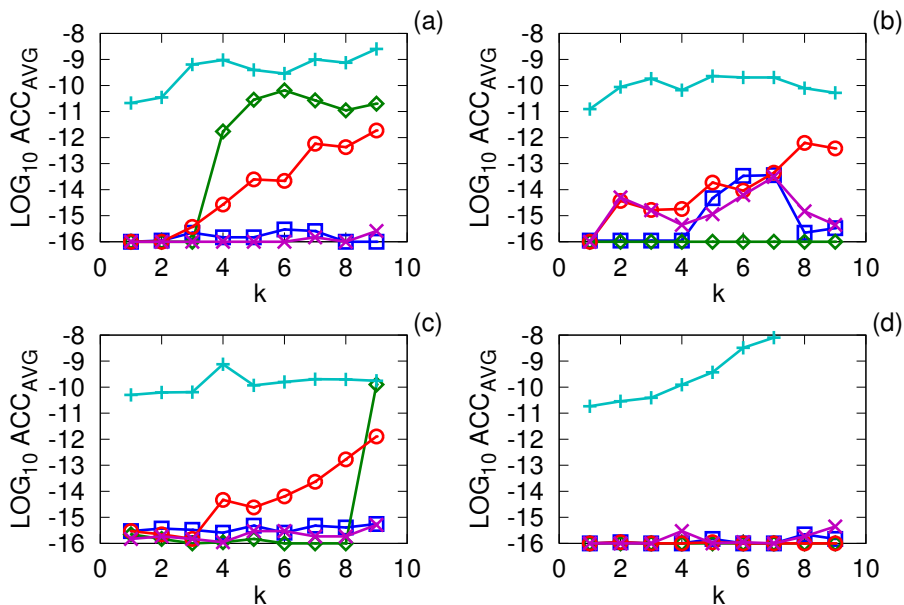}}
\caption{Accuracy performance for $N_D =  2$.  Routines are indicated by $\Box$ for DBLGK, $\Diamond$ for DBLCC, $\bigcirc$ for AMGKQ, and $+$ for ADAPT.  The integrands are normal distributions in (a), Cauchy distributions in (b), beta distributions in (c), and sinusoids in (d).  Results indicated by $\times$ are the accuracy when AMGKQ does all four integrands simultaneously.}
\label{fig:D}
\end{figure}

We can see in Figure~\ref{fig:D} that the accuracy of AMGKQ is comparable to that of DBLGK and DBLCC.  Upon termination, all three produce a result whose accuracy is well below the requested precision.  In contrast, ADAPT returns a result whose accuracy is on par with $E_A$; to be fair, that is all we asked for, but as we will see next, ADAPT has to work much harder to achieve a result that is nowhere near as accurate as the others.  While \citet{Berntsen-1991437} recommend using a higher order rule for oscillatory integrands, the default rule of order 7 is used in ADAPT when evaluating the product of sinusoids in panel (d) for consistency of comparison.  The relative performance of DBLGK and DBLCC varies with the choice of integrand, whereas AMGKQ is more consistent in that regard.  Also shown is the accuracy of AMGKQ when it does all four integrands simultaneously.

\begin{figure}
\centerline{\includegraphics[scale=.8]{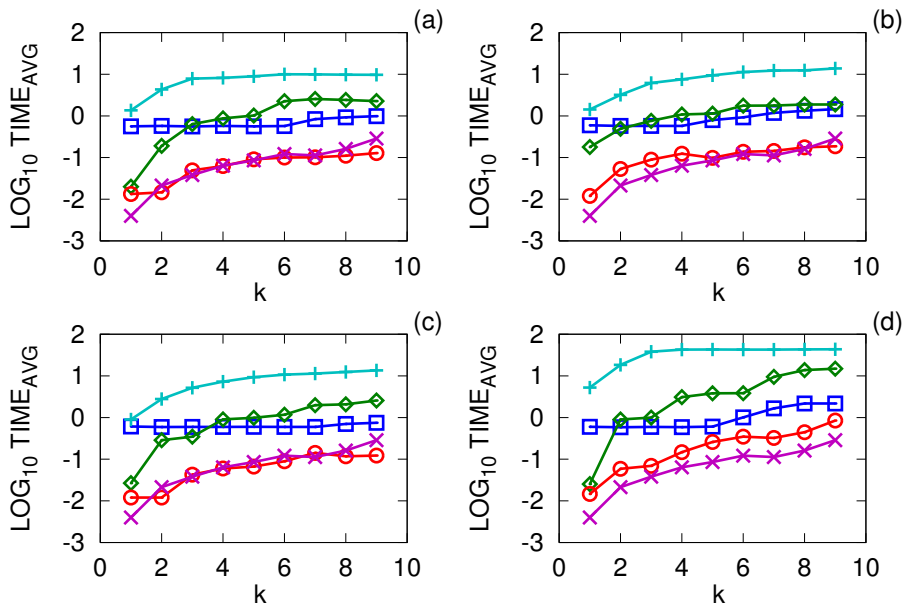}}
\caption{Running time performance for $N_D =  2$.  Routines are indicated by $\Box$ for DBLGK, $\Diamond$ for DBLCC, $\bigcirc$ for AMGKQ, and $+$ for ADAPT.  The integrands are normal distributions in (a), Cauchy distributions in (b), beta distributions in (c), and sinusoids in (d).  Results indicated by $\times$ are one fourth of the running time when AMGKQ does all four integrands simultaneously.}
\label{fig:E}
\end{figure}

In Figure~\ref{fig:E} we compare the running times of the various implementations.  Since the integrand functions are called in slightly different ways between DBLQUAD and AMGKQ, it is not really fair to compare the number of calls.  The practical quantity which the user wants to minimize is running time, which is measured here in terms of CPU seconds, not wall time.  For easy integrands (small $k$), AMGKQ returns a result up to 100 times faster than DBLGK, while for more difficult integrands the speedup factor is closer to 10, and it does not slow down as much as does DBLCC with increasing $k$.  A factor of 10 might not seem like much for an operation that takes only a second, but in the context of Bayesian data analysis one often has to repeat variations of the same integral a large number of times.  Furthermore, for real time predictive applications, every CPU cycle counts.

We should mention that the initialization time for AMGKQ is not included in these comparisons.  If it were, it would only affect $k=1$ in panel (a) of Figure~\ref{fig:E}, raising its value by a factor of 10.  After the first call, which need not produce anything useful, all the machinery for subsequent calls at the same order $n_G$ in the same number of dimensions $N_D$ is available in memory; since that feature is not part of the other implementations, it does not make sense to penalize AMGKQ for its inclusion.  At any rate, all four integrands for each $k$ can be evaluated by AMGKQ in one pass, thus doing them independently is already generous to its competition.  The one pass running time displayed in the figure is one quarter of the time to do all four integrands simultaneously.

\begin{figure}
\centerline{\includegraphics[scale=.8]{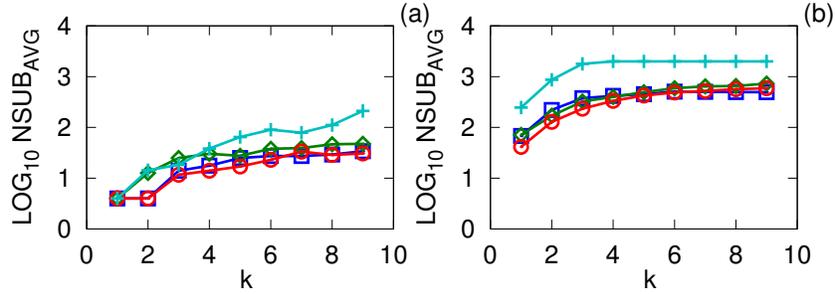}}
\caption{Comparison of the number of subregions evaluated $N_s$ for AMGKQ in panel (a) and ADAPT in panel (b).  The integrands are indicated by $\Box$ for the normal distributions, $\Diamond$ for the Cauchy distributions, $\bigcirc$ for the beta distributions, and $+$ for the sinusoids.}
\label{fig:F}
\end{figure}

Let us next compare the number of subregions evaluated by AMGKQ and ADAPT, as shown in Figure~\ref{fig:F}.  In panel (a) one sees that for the smallest $k$, AMGKQ converges after the initial subregions have been evaluated, while ADAPT requires on the order of 100 subregions (iterations) before it converges.  For the larger $k$, there is a modest increase in $N_s$ upon termination for AMGKQ, while ADAPT reaches its maximum limit of $N_S = 2000$ for the sinusoidal integrand.  The vastly different values of $N_s$ for these two algorithms can only be explained by the superior performance of the Gauss-Kronrod quadrature rules.

\begin{figure}
\centerline{\includegraphics[scale=.8]{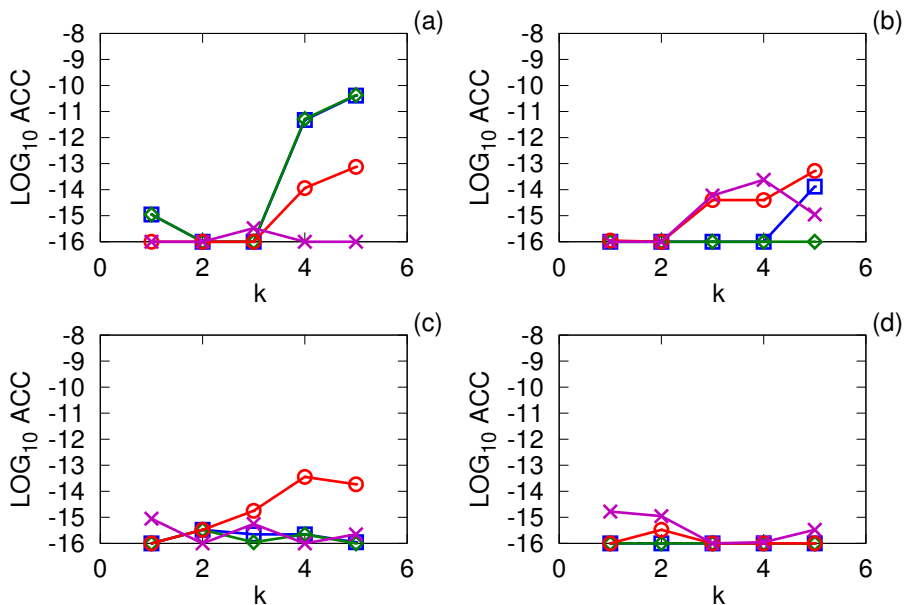}}
\caption{Accuracy performance for $N_D =  3$.  Routines are indicated by $\Box$ for TPLGK, $\Diamond$ for TPLCC, and $\bigcirc$ for AMGKQ.  The integrands are normal distributions in (a), Cauchy distributions in (b), beta distributions in (c), and sinusoids in (d).  Results indicated by $\times$ are the accuracy when AMGKQ does all four integrands simultaneously.}
\label{fig:G}
\end{figure}

\begin{figure}
\centerline{\includegraphics[scale=.8]{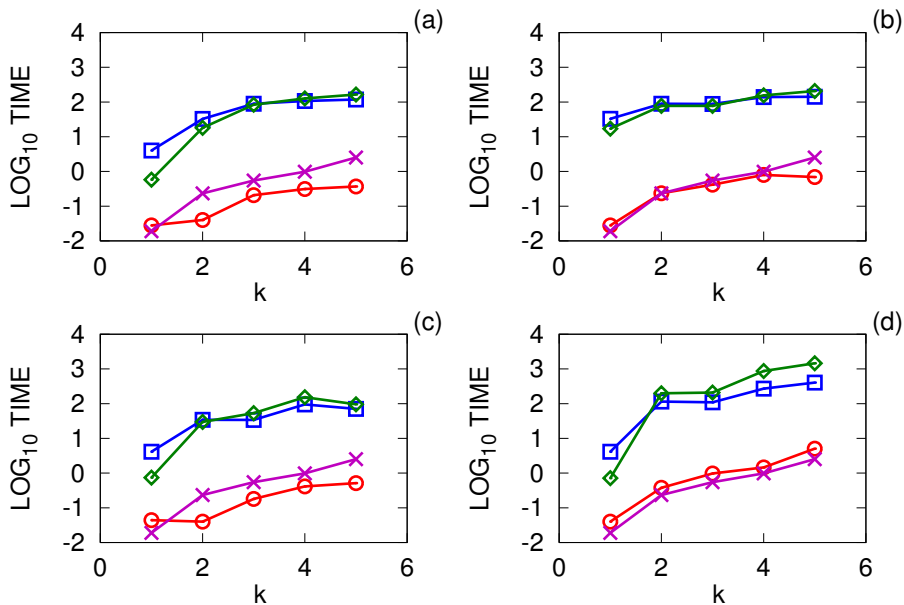}}
\caption{Running time performance for $N_D =  3$.  Routines are indicated by $\Box$ for TPLGK, $\Diamond$ for TPLCC, and $\bigcirc$ for AMGKQ.  The integrands are normal distributions in (a), Cauchy distributions in (b), beta distributions in (c), and sinusoids in (d).  Results indicated by $\times$ are one fourth of the running time when AMGKQ does all four integrands simultaneously.}
\label{fig:H}
\end{figure}

We can repeat the comparison for $N_D = 3$, at least for small values of $k$.  Only a single set of runs is considered, on account of the length of time TRIPLEQUAD takes to converge.  Likewise, ADAPT is no longer considered for the same reason.  The accuracies shown in Figure~\ref{fig:G} of TPLGK, TPLCC, and AMGKQ are all, as expected, well below the requested precision, as is the accuracy of the simultaneous integrands.  What is interesting is the comparison of their running times, displayed in Figure~\ref{fig:H}.  When the integrals are done independently, we see that AMGKQ outperforms TRIPLEQUAD by a factor greater than 100 and sometimes close to 1000.  Interestingly, doing the integrals simultaneously appears to take slightly longer than their aggregate time, but the accuracy is not allowed to drift as much for large $k$.  Obviously, performing multivariate quadrature with a recursive algorithm is not the quickest path to success.

\section{Environment Limitations and the Curse of Dimensionality}

As implemented, AMGKQ is not self-limiting; the number of dimensions $N_D$, the number of integrands $N_F$, and the order of quadrature rules $(n_G, n_K)$ can be arbitrarily large.  Of course, there are practical limits imposed by the operating environment, which is comprised of the hardware and the interpreter.  The two largest objects which AMGKQ holds in memory are the abscissa locations in physical units $\msf{X}_s$ of size $[N_D, N_X]$ and the corresponding integrand values $\msf{Y}_s$ of size $[N_F, N_X]$, where $N_X$ is determined by the quadrature order $(n_G, n_K)$ and the number of dimensions $N_D$.  When variable transformations are in play, there can be a succession of functions that each create an array the size of $\msf{X}_s$ in memory, and there must be room in RAM to hold them.

Another limit is imposed by the class of indexing variable, which is implemented in Octave as a signed integer.  On a 32-bit system, the maximum number of elements $N_{32}$ that can be stored in an array is one less than the maximum positive integer that can be represented, or $N_{32} = 2^{31} - 2$.  If either product $N_D N_X$ or $N_F N_X$ is greater than $N_{32}$, AMGKQ will fail to allocate room in memory for $\msf{X}_s$ or $\msf{Y}_s$, respectively.  On a fully 64-bit system (hardware and interpreter), the number of elements possible $N_{64}$ is much greater.  The relation between the number of abscissae and the order of quadrature is $N_X = (2 n_G + 1)^{N_D}$, such that a greater number of dimensions is feasible if one reduces the quadrature order.

\begin{table}[]\footnotesize
\caption{Multivariate Tests At Order $n_G = 7$\label{tab:D}}
\centering
\begin{tabular}{ccc|c|cc|cc}
\hline
$ N_D $ & $ N_X $ & $ N_D N_X $ & Nos. & $N_s$ & flag & ERR & ACC \\\hline 
1 & 15 & 15 & 1 & 2 & 2 & $ \begin{tabular}{c} 1.1e-16 \end{tabular} $ & $ \begin{tabular}{c} 2.2e-16 \end{tabular} $ \\ 
2 & 225 & 450 & 1,2 & 4 & 2 & $ \begin{tabular}{c} 9.7e-13 \end{tabular} $ & $ \begin{tabular}{c} 2.2e-16 \end{tabular} $ \\ 
3 & 3375 & 10125 & 1,2,3 & 8 & 2 & $ \begin{tabular}{c} 2.6e-13 \end{tabular} $ & $ \begin{tabular}{c} 1.1e-16 \end{tabular} $ \\ 
4 & 50625 & 202500 & 1,2,3,4 & 16 & 2 & $ \begin{tabular}{c} 1.5e-13 \end{tabular} $ & $ \begin{tabular}{c} 5.6e-16 \end{tabular} $ \\ 
5 & 759375 & 3796875 & 1,2,3,4,5 & 32 & 2 & $ \begin{tabular}{c} 1.9e-14 \end{tabular} $ & $ \begin{tabular}{c} 1.1e-16 \end{tabular} $ \\ 
6 & 11390625 & 68343750 & 1,2,3,4,5,6 & 64 & 2 & $ \begin{tabular}{c} 6.4e-15 \end{tabular} $ & $ \begin{tabular}{c} 6.9e-17 \end{tabular} $ \\ 

\hline
\end{tabular}
\end{table}%

\begin{figure}[]
\centerline{\includegraphics[scale=.8]{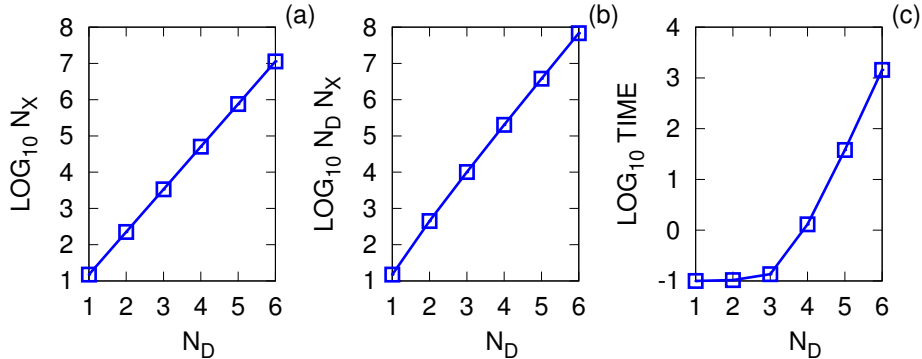}}
\caption{The curse of dimensionality.}
\label{fig:I}
\end{figure}

There is simply no getting around the fact that the accurate numeric approximation of the integral of a multivariate function requires a lot of work.  As a final test of AMGKQ, let us evaluate at order $n_G = 7$ an integrand comprised of the product of the first few Burkardt tests in one dimension, with each additional function evaluated from an independent variable.  As we can see in Table~\ref{tab:D} and Figure~\ref{fig:I}, the number of elements in $\msf{X}_s$ and $\msf{Y}_s$ grows quite quickly with dimension.  The running time shown in panel (c) includes the initialization time, since we are comparing apples to apples here, and demonstrates the curse of dimensionality.  Not only must AMGKQ evaluate a strongly growing number of elements in $\msf{X}_s$ and $\msf{Y}_s$ as $N_D$ increases, but it also must do more work to evaluate each dimension's contribution to the integrand.  Inspecting the column for $N_s$ in the table, we see that AMGKQ converged for all these integrals immediately after initialization.  When the sixth function was appended, the Octave process consumed over 2GB of RAM and took more than a few minutes of wall time to evaluate.  On a modern platform, the algorithm should be able to handle more dimensions than are considered here.

\section{Outlook}

During final preparations, the algorithm CHEBINT \citep{Poppe-2013401} has come to our attention.  This work focuses on the exposition of AMGKQ and its comparison to its parent algorithms ADAPT and QUADGK.  It would be interesting in future work to compare the performance of AMGKQ and CHEBINT directly.  Efficiency can be measured not only in terms of running time but also in terms of code complexity; AMGKQ accomplishes its goals with less than 1000 lines of code, including comments, examples, and nearly 200 lines of tabulated coefficients.

There remain opportunities to improve the efficiency of the implementation of AMGKQ.  The most obvious upon reading the code is the manner in which the variable transformations are addressed.  Rather than performing the transformations sequentially, it would be better to identify for each dimension the required transformation(s) and then effect the change of variable in a single function to reduce memory overhead and other expenses.  It might also be better to transpose the sense in which $\msf{X}_s$ and $\msf{Y}_s$ are stored.  Such detailed investigations of efficiency improvement are left for the interested reader to perform.

\section{Conclusion}

This work describes an efficient algorithm for the adaptive multivariate Gauss-Kronrod quadrature of simultaneous integrands and its implementation in Octave, AMGKQ.  Its accuracy is comparable to the numerical integration routines provided by Octave, and its running time is much faster in multiple dimensions.  Its performance is achieved by using vectorized code as much as possible, including in the user supplied integrand function.  Its performance is limited only by the memory structure of its operating environment.  The numeric approximation of integrals of functions of several variables might not be easy, but it has at least become easier.


\section*{Acknowledgement}
The author would like to thank Dr. Yonggang Liu of the University of South Florida for assisting with the compatibility testing.

\bibliographystyle{plainnat}


\end{document}